\documentstyle[aps,graphicx]{revtex}

\tightenlines
\twocolumn
\begin{document}
\draft
\preprint{ITAMP Preprint 99/11}
\title{
Role of negative-energy states and Breit interaction in
calculation of atomic parity-nonconserving amplitudes.
}

\author{A. Derevianko} 
\address{Institute for Theoretical Atomic and Molecular Physics\\
Harvard-Smithsonian Center for Astrophysics,
Cambridge, Massachusetts 02138}

\date{\today}
\maketitle

\begin{abstract}
It is demonstrated that Breit and negative-energy state contributions
reduce the  $2.5\sigma$ deviation [S.C. Bennett and C.E. Wieman,
Phys.\ Rev.\ Lett.\ {\bf 82}, 2484 (1999)] in the value of the weak 
charge of $^{133}$Cs  from the Standard Model prediction to 1.7$\sigma$.  
The corrections are
obtained in the relativistic many-body perturbation theory by combining
all-order Coulomb and second-order Breit contributions.
The corrections to parity-nonconserving  amplitudes 
amount to 0.6\% in $^{133}$Cs and 1.1\% in $^{223}$Fr.
The relevant magnetic-dipole 
hyperfine structure constants are modified  
at the level of 0.3\% in Cs, and 0.6\% in Fr.
Electric-dipole matrix elements are affected at  0.1\% level in Cs and 
a few 0.1\% in Fr.
\end{abstract}


Atomic parity-nonconserving (PNC) experiments combined
with accurate atomic structure calculations provide  constrains
on ``new physics'' beyond the Standard Model of elementary-particle
physics. 
Compared
to high-energy experiments or low-energy $ep$ scattering experiments, 
atomic single-isotope PNC measurements are uniquely sensitive
to new isovector heavy physics~\cite{Ramsey-Musolf_PRC99}.
Presently, the PNC effect in atoms was most precisely measured 
by Boulder group in  $^{133}\mathrm{Cs}$~\cite{Wood_PNC97}.
They determined ratio of PNC amplitude $E_{\mathrm{ PNC}}$ to
the tensor transition polarizability $\beta$ for $7S_{1/2}-6S_{1/2}$ transition
with a precision of 0.35\%. 
In 1999, Bennett and Wieman~\cite{BennettWieman_PRL99} 
accurately measured
tensor transition polarizability $\beta$, and by combining the
previous theoretical determinations of the $E_{\mathrm{ PNC}}$~\cite{Blundell_PNC,Dzuba_PL89} 
with their measurements, they 
have found a value of the weak charge for 
$^{133}\mathrm{Cs}$  
$Q_{\mathrm W} = - 72.06(28)_{\mathrm{expt}} (34)_{\mathrm{theor}}$
which differed from the prediction~\cite{QW_StdModel} 
of the Standard Model 
$Q_{\mathrm W} = - 73.20(13)$
by 2.5 standard deviations. They also reevaluated the precision 
of the early 1990s atomic structure calculations~\cite{Blundell_PNC,Dzuba_PL89},
and argued that the uncertainty of the predicted  $E_{\mathrm{ PNC}}$  is 0.4\%,
rather than previously estimated 1\%. This conclusion has been  
based on a much better agreement of calculated and recently
accurately measured electric-dipole amplitudes for the resonant transitions in alkali-metal atoms.

In view of the reduced uncertainty, the purpose of this Letter is
to evaluate contributions from negative-energy states (NES) and
Breit interaction. It will be demonstrated that these
contributions correct  theoretical $E_{\mathrm{ PNC}}$ 
and the resultant value of the weak charge by 0.6\% in Cs.
It is worth noting, that due to the smallness of these contributions at 
what had been believed to be a 1\%  theoretical error in Cs, 
the previous calculations have either omitted~\cite{Dzuba_PL89},  
or estimated the contributions  from Breit interaction only partially~\cite{Blundell_PNC}.
The main focus of the previous {\em ab initio} 
calculations has been correlation contribution from the residual Coulomb interaction (i.e. beyond
Dirac-Hartree-Fock level). In both calculations important chains of many-body diagrams have been summed
to all orders of perturbation theory.

This Letter also reports correction due to NES and Breit interaction for $E_{\rm PNC}$ in francium.
The interest in Fr stems from the fact that 
analogous  PNC amplitude is 18 times larger in heavier $^{223}$Fr compared to Cs~\cite{Dzuba_FrPNC}.
The measurement of atomic PNC in Fr is pursued by Stony Brook 
group~\cite{PNC_prospects_Fr}.

The quality of theoretical atomic wave-functions at small radii  
is usually judged by comparing calculated and experimental 
hyperfine-structure magnetic-dipole constants $A$, and
in the intermediate region by comparing electric-dipole matrix elements. 
It will be demonstrated that the corresponding corrections 
to all-order Coulomb values are at the level of a few 0.1\%.

The PNC amplitude  of 
$nS_{1/2} \rightarrow n'S_{1/2}$ transition can be represented as a sum
over intermediate states $mP_{1/2}$
\begin{eqnarray}
\lefteqn{E_{\mathrm PNC} = \sum_{m} 
\frac{\langle n'S|D|mP_{1/2}\rangle  \langle mP_{1/2} |H_{W}|nS\rangle
}{E_{nS}-E_{mP_{1/2}}}  } \nonumber \\  &+ &
\sum_{m} 
\frac{\langle n'S|H_{W}|mP_{1/2}\rangle  \langle mP_{1/2} |D|nS\rangle
}{E_{n'S}-E_{mP_{1/2}}}  
\, .
\label{Eqn_E_PNC}
\end{eqnarray}
The overwhelming contribution from parity-violating interactions arises from 
the Hamiltonian
\begin{equation}
 H_{\rm W} = \frac{G_F}{\sqrt{8}} Q_{\rm W} \rho_{\mathrm nuc}(r) \gamma_5 \, ,
\label{Eqn_Hw}
\end{equation}
where $G_F$ is the Fermi constant,$\gamma_5$ is the Dirac matrix,
and $\rho_{\mathrm nuc}(r)$ is the nuclear distribution.
To be consistent with the previous calculations
the $\rho_{\mathrm nuc}(r)$  is taken to be a Fermi distribution with the ``skin depth'' 
$a=2.3/(4\ln 3)$ fm
and the cutoff radius $c=5.6743$ fm for  $^{133}\mathrm{Cs}$ as in Ref.~\cite{Blundell_PNC},
and $c=6.671$ fm for $^{223}$Fr as in Ref.~\cite{Dzuba_FrPNC}.
The PNC amplitude is customarily expressed in the units of $10^{-11} i (-Q_{\rm W}/N)$,
where  $N$ is the number of neutrons in the nucleus
($N=78$ for $^{133}$Cs and $N=136$ for $^{223}$Fr ). Atomic units  are 
used throughout the Letter. 
The results of the calculations
for  $^{133}\mathrm{Cs}$ are  
$E_{\mathrm PNC} = -0.905   \times 10^{-11} i(-Q_{\rm W}/N)$, Ref.~\cite{Blundell_PNC} and
$E_{\mathrm PNC} = -0.908  \times 10^{-11} i(-Q_{\rm W}/N)$, Ref.~\cite{Dzuba_PL89}.
The former value includes a partial Breit contribution $+0.002 \times 10^{-11} i(-Q_{\rm W}/N)$, 
and the latter does not. Both calculations are in a very close agreement if the Breit contribution
is added to the value of Novosibirsk group. The reference many-body Coulomb value
\begin{equation}
 E^{C}_{\mathrm PNC} = -0.9075   \times 10^{-11} i(-Q_{\rm W}/N)
\label{Eqn_refVal}
\end{equation}
is determined as an average of the two, with Breit contribution removed from the 
value of Notre Dame group.
For $^{223}$Fr $E_{\mathrm PNC} = 15.9 \times 10^{-11} i (-Q_{\rm W}/N)$, Ref.~\cite{Dzuba_FrPNC},
this value  does not include Breit interaction.

{\em Ab initio} relativistic many-body calculations of wave-functions, like
coupled-cluster type calculations of~\cite{Blundell_PNC,Safronova_AlkSDpT99}, 
to avoid the ``continuum dissolution problem''~\cite{Sucher},
start from the {\em no-pair} Hamiltonian derived 
from QED~\cite{BrownRavenhall51}. 
The {\em no-pair} Hamiltonian
excludes
virtual electron-positron pairs from the resulting correlated wave-function. 
If the {\em no-pair} wave-functions are further used to obtain
many-body matrix elements, the negative-energy state (NES) contribution is
missing already in the second order. Recently, it has been shown
that the magnetic-dipole transition amplitude both in He-like ions~\cite{Derevianko_NESHe98} 
and alkali-metal atoms~\cite{Savukov_PRL99}
can be strongly affected by the NES correction.  The
enhancement mechanism is  due to  vanishingly  small 
lowest-order values and also due to mixing of  large and small components of a Dirac wavefunction
by  magnetic-dipole operator. For the NES ($E< -mc^2$) the meaning of large and
small components is reversed, i.e., small component
is much larger than large component. The mixing of large positive-energy
component with small component of NES  results in much larger one-particle
matrix elements, than in the {\em no-pair} case. 
The $2mc^2$ energy denominators lessen the effect, but, for example,
the Rb $5S_{1/2} - 6S_{1/2}$
magnetic-dipole rate is reduced by a factor of 8 from the {\em no-pair} value
by the inclusion of NES~\cite{Savukov_PRL99}.  The inclusion of  
Breit interaction also becomes important and the size of the correction is 
comparable to the Coulomb contribution.
Just as
in the case of magnetic-dipole operator, the Dirac matrix $\gamma_5$  in the weak 
Hamiltonian Eq.~(\ref{Eqn_Hw}) mixes
large and small components of  wavefunctions. Similar mixing occurs in the
matrix element describing interaction of an electron with nuclear magnetic moment 
(hyperfine structure constant $A$). As demonstrated below, the relative effect for
these operators is not 
as strong as in the magnetic-dipole transition case, since 
the lowest order matrix elements are nonzero in the nonrelativistic limit, but is still 
important.
It is worth noting that the problem of NES does not appear explicitly in the
Green's function~\cite{Dzuba_PL89} or mixed-parity~\cite{Blundell_PNC} 
approaches; 
however the correction due to Breit interaction 
still has to be addressed. The NES Coulomb corrections have to be taken into account
explicitly in the ``sum-over-states'' method~\cite{Blundell_PNC}, employing
all-order many-body values obtained with the {\em no-pair} Hamiltonian. 

The analysis is based on the $V_{N-1}$ Dirac-Hartree-Fock potential wave-functions,
with the valence and virtual orbitals $m$ calculated in the ``frozen'' potential
of core orbitals $a$.  
The second-order correction to a matrix element of one-particle operator $Z$ between two valence
states $w$ and $v$ is represented as 
\begin{eqnarray}
Z_{wv}^{(2)} &=&\sum\limits_{i\neq v}\frac{z_{wi\,}b_{iv}}{\epsilon
_{v}-\epsilon _{i}}+\sum\limits_{i\neq w}\frac{b_{wi}\,z_{iv}}{\epsilon
_{w}-\epsilon _{i}}+  \nonumber\\
+\sum\limits_{ma} && \! \! \frac{z_{am}(\widetilde{g}_{wmva}+\widetilde{b}_{wmva})}{%
\epsilon _{a}+\epsilon _{v}-\epsilon _{m}-\epsilon _{w}}
+\sum\limits_{ma}
\frac{(\widetilde{g}_{wavm}+\widetilde{b}_{wavm})\,z_{ma}}{\epsilon
_{a}+\epsilon _{w}-\epsilon _{m}-\epsilon _{v}} \, . \label{EqnZ2MBPT}
\end{eqnarray}
This expression takes into account the 
residual (two-body) Coulomb, $g_{ijkl}$, and
two-body $b_{ijkl}$ and one-body $b_{ij}=\sum_{a} \tilde{b}_{iaja}$ Breit interaction.
Static form of Breit interaction
is used in this work.
The tilde denotes antisymmetric combination $\tilde{b}_{ijkl} = b_{ijkl}-b_{ijlk}$.
Subscript $i$ ranges over both core and excited states. 
Note that summation over states $i$ and $m$
includes negative-energy states.   
The NES correction to PNC amplitudes arises in two circumstances,
directly from the sum in Eq.~(\ref{Eqn_E_PNC}) and in the values of electric-dipole and weak interaction
matrix elements.
If the length-gauge of the electric-dipole operator is
 used, the direct contribution of NES in the amplitude Eq.~(\ref{Eqn_E_PNC}) is 
a factor of $10^{-13}$ smaller than the total amplitude, and will be disregarded in the
following. The numerical summations are done using 100 positive- and 100 negative-energy
wavefunctions in a B-spline representation~\cite{Bsplines} 
obtained in a cavity  with a radius of  75 a.u.

The breakdown of second-order corrections to matrix elements of weak interaction
for Cs is given in Table~\ref{Tbl_Cs_weak}. The all-order values from Ref.~\cite{Blundell_PNC}
are also listed in the table to fix the relative phase of the contributions. 
The matrix elements are each modified at 0.6-0.7\% level. Most of the correction
arises from positive-energy Breit contribution, negative-energy states contribute
at a smaller but comparable level. The contributions  from NES due to one-body and two-body
Breit interaction are almost equal ( $B^{(1)}_- \approx B^{(2)}_-$) and, in addition,
$B^{(1)}_+ \approx 2 B^{(2)}_+$. The same relations hold also in francium. 
The corrections to the relevant length-form matrix elements are overwhelmingly
due to the one-body Breit interaction, and are at 0.1\% level. For example,
the all-order {\em reduced} matrix element 
$\langle 6S_{1/2} || D || 6P_{1/2} \rangle=4.478$ Ref.~\cite{Safronova_AlkSDpT99}
is increased by 0.005, bringing the total 4.483 into an
excellent agreement with experimental value~\cite{CsE1} 4.4890(65). 
Generally, the corrections reduce absolute values of the weak interaction
matrix elements, and increase absolute values of the dipole matrix elements, 
therefore, 
their net contributions to $E_{\rm PNC}$ have an opposite sign. Matrix elements of weak
interaction are affected more strongly, because of the sampling of wave-function in the 
nucleus, where relativity is important.  

As demonstrated in Ref.~\cite{Blundell_PNC}, the four lowest-energy valence $mP_{1/2}$ states contribute 98\% of the sum in Eq.~(\ref{Eqn_E_PNC}),
and for the purposes of this work, limiting the sums to only these states is sufficient.
The corrections to  $E_{\rm PNC}$ are calculated first by replacing 
the weak interaction matrix elements
with the relevant second-order contributions and at the same time using all-order
dipole matrix elements, and second by taking all-order $H_W$ matrix elements, and replacing $D$ with 
the appropriate correction. 
In both cases the experimental energies are used in the denominators. 
The needed all-order matrix elements for Cs are tabulated in Ref.~\cite{Blundell_PNC}.  
The summary of corrections to  $E_{\rm PNC}$ is presented in Table~\ref{Tab_Epnc}. The modifications
in the weak interaction matrix element provide a dominant correction. The contribution
due to NES in the Coulomb part is insignificant, and is already effectively
included in the reference many-body Coulomb
value $E^{C}_{\mathrm PNC}$,  Eq.~(\ref{Eqn_refVal}).
The reference value  $E^{C}_{\mathrm PNC}$ is modified by the Breit contributions by 0.6\%, almost
two times larger than the uncertainty in the Boulder experiment~\cite{Wood_PNC97}.
The modified value is
\[
 E^{C+B}_{\mathrm PNC}(^{133}{\rm Cs} ) = -0.902(36) \times 10^{-11} i(-Q_{\rm W}/N) \, .
\]
A 0.4\% uncertainty had been assigned to the above result following analysis of 
Bennett and Wieman~\cite{BennettWieman_PRL99}. 
When  $E^{C+B}_{\mathrm PNC}$ is  combined with  the experimental values of transition 
polarizability $\beta$~\cite{BennettWieman_PRL99} and $E_{\rm PNC}/\beta$~\cite{Wood_PNC97},
one obtains for the weak charge
\[
Q_{\rm W}(^{133}{\rm Cs} ) = -72.42(28)_{\rm expt}(34)_{\rm theor} \, .
\]
This value differs from the prediction of the Standard Model $Q_{\rm W}=-73.20(13)$ by 1.7$\sigma$,
versus 2.5$\sigma$ discussed in Ref.~\cite{BennettWieman_PRL99}, where $\sigma$ is calculated
by taking uncertainties in quadrature.

The only previous calculation of Breit contribution to PNC amplitude in Cs
has been performed by the Notre Dame group~\cite{Blundell_PNC}, using mixed-parity Dirac-Hartree-Fock
formalism. The one-body Breit interaction 
has been included on equal footing with the DHF potential,
but the linearized modification to one-body Breit potential  
due to $H_{\rm W}$ ($V_{\rm PNC-HFB}$ in notation of Ref.~\cite{Blundell_PNC}) has been omitted.
It is straightforward to demonstrate that because of this omission,  
the comparable contribution from
{\em two-body} part of the Breit interaction has been disregarded.
In units of  $10^{-11} i (-Q_{\rm W}/N)$, the result of the present calculation for
{\em one-body} Breit contribution is 0.003 versus 0.002
in Ref.~\cite{Blundell_PNC}. Such disagreement is most probably caused by different
types of correlation contribution included in the two approaches.
Treating one-body Breit together with the DHF potential 
effectively sums the many-body contributions from one-body Breit interaction
to all orders, and presents the advantage of the scheme employed in Ref.~\cite{Blundell_PNC}. 
However, the dipole matrix elements and energies in the sum Eq.~(\ref{Eqn_E_PNC}) 
are effectively included at the  DHF level in the formulation of Ref.~\cite{Blundell_PNC},
in contrast to high-precision 
all-order values employed  in the present work.
The difference between the two values can be considered as
a theoretical uncertainty in the value of the Breit correction. Clearly
more work needs to be done to resolve the discrepancy. 
The accuracy of the present
analysis can be improved if the {\em one-body} Breit interaction is embodied in DHF equations, and the many-body
formulation starts from the resulting basis. However, to improve present second-order
treatment of the {\em two-body} part of the Breit interaction, higher orders of perturbation theory 
have to be considered.
Apparently the most important contribution would arise from terms linearized 
in the Breit interaction, i.e. diagrams
containing one matrix elements of the Breit interaction and the rest of  
the residual Coulomb interaction. 

The Breit and NES corrections to PNC amplitude in heavier Fr are more pronounced.
The $^{223}$Fr PNC amplitude 15.9 $\times 10^{-11} i (-Q_{\rm W}/N)$  from Ref.~\cite{Dzuba_FrPNC} 
is reduced by 1.1\%.
Using all-order dipole matrix elements from Ref.~\cite{Safronova_AlkSDpT99},
the following corrections due to modifications in the $h_W$ are found 
(in units of $10^{-11} i (-Q_{\rm W}/N)$):
one-body Breit $B^{(1)}_{\pm} = -0.131$,  two-body Breit $B^{(2)}_{\pm}= -0.053$,
and the $C_-$ correction, implicitly included in Ref.~\cite{Dzuba_FrPNC},
is $-0.003$. As in the case of Cs, the all-order {\em no-pair}
Coulomb result~\cite{Safronova_AlkSDpT99} for reduced
matrix element $\langle 7P_{1/2} || D||7S_{1/2}\rangle = 4.256$ is increased
by inclusion of the Breit interaction and NES by $0.0011$, a 0.3\% modification, leading 
to a much better agreement with experimental value 4.277(8)~\cite{FrE1}.
The modification of the $E_{\mathrm PNC}$ due to corrections
in the dipole matrix elements is much smaller 
than in the case of $h_{\rm W}$. 
At present there is no tabulation of accurate
matrix elements of weak interaction for Fr, and the influence on $E_{\rm PNC}$ due to
the Breit contribution in dipole matrix elements is
estimated  from average of the modification of individual dipole matrix elements 0.2\%.
The {\em net } result decreases 
the reference Coulomb value for $^{223}$Fr~\cite{Dzuba_FrPNC} by 0.18 $\times 10^{-11} i (-Q_{\rm W}/N)$,
and the corrected value is
\[
 E^{C+B}_{\mathrm PNC}(^{223}{\rm Fr} ) = 15.7 \times 10^{-11} i(-Q_{\rm W}/N) \, .
\]

Finally, it is worth discussing Breit and NES contributions to 
hyperfine-structure magnetic-dipole constants $A$ for the states involved 
into PNC calculations.
The all-order {\em no-pair} Coulomb values in the recent work~\cite{Safronova_AlkSDpT99} 
have been corrected using a similar second-order formulation;  
no details of the calculation have been given. The explicit contributions
listed in Table~\ref{Tbl_hfs} will be useful for correcting {\em ab initio}
many-body Coulomb values. The table presents
the contributions for two lowest valence
$S_{1/2}$ and $P_{1/2}$ states. The calculations are performed
using a model of uniformly magnetized nucleus with a magnetization radius $R_m$ given
in the table. 
One finds that the additional terms reduce values
calculated in the {\em no-pair} Coulomb-correlated approach.
For Cs the corrections are of order 
0.2\% for $6S_{1/2}$, 0.1\% for $7S_{1/2}$, and 
0.3\% for  $6P_{1/2}$ and $7P_{1/2}$. The  
relative contributions to hyperfine constants in heavier Fr are larger,
accounting for 0.5\% of the total value for $7S_{1/2}$, 0.4\% for $8S_{1/2}$,
and 0.6\% for $7P_{1/2}$ and $8P_{1/2}$.

This work
demonstrates that the Breit and NES contributions are
comparable to the remainder of Coulomb correlation corrections
unaccounted for in  modern relativistic all-order 
many-body calculations and hence have to be systematically 
taken into account.
In particular, the Breit interaction contributes 0.6\% to parity-nonconserving
amplitudes in Cs and 1.1\% in Fr. 
The correction for Cs is almost twice the experimental
uncertainty and reduces the recently determined~\cite{BennettWieman_PRL99} 2.5 $\sigma$ deviation 
in the value of weak charge from the Standard Model prediction to 1.7$\sigma$.
Both hyperfine constants and electric-dipole
matrix elements  are affected at a few 0.1\%. By including NES and Breit correction,
the {\em no-pair} Coulomb all-order dipole matrix
elements~\cite{Safronova_AlkSDpT99} for resonant transitions are brought 
into an excellent agreement with the
accurate experimental values.

This work was supported by the U.S. Department of Energy,
Division of Chemical Sciences, Office of Energy Research. 
The author would like to thank W.R. Johnson for useful discussions
and H.R. Sadeghpour for suggestions on manuscript.

\begin{table}
\caption{ Contributions to matrix elements of the weak interaction for 
$^{133}$Cs in units $10^{-11} i (-Q_{\rm W}/N)$. 
All-order {\em no-pair}  values are from Blundell {\em et al.}
\protect\cite{Blundell_PNC}. $C_{-}$ is the correction from negative-energy states for
the residual Coulomb interaction, and $B^{(1)}_{\pm}$ and $B^{(2)}_{\pm}$ are positive/negative energy 
state contributions  from one-body and two-body Breit interaction. Notation $x[y] = x \times 10^{y}$.
\label{Tbl_Cs_weak}}
\begin{tabular}{lrrrrrr}
\multicolumn{1}{c}{$n$} & 
\multicolumn{1}{c}{ all-order} & 
\multicolumn{1}{c}{$C_-$} & 
\multicolumn{1}{c}{$B^{(1)}_+$} & 
\multicolumn{1}{c}{$B^{(1)}_-$} & 
\multicolumn{1}{c}{$B^{(2)}_+$} & 
\multicolumn{1}{c}{$B^{(2)}_-$} \\
\hline
\multicolumn{1}{c}{} &
\multicolumn{6}{c}{$\delta \langle nP_{1/2}  |h_W| 6S_{1/2} \rangle$} \\
 6&  5.62[-2]&   -9.05[-6]& -3.10[-4]&  5.64[-5]& -1.54[-4]&  5.73[-5]\\
 7&  3.19[-2]&   -5.41[-6]& -1.82[-4]&  3.37[-5]& -9.21[-5]&  3.43[-5]\\
 8&  2.15[-2]&   -3.71[-6]& -1.23[-4]&  2.31[-5]& -6.32[-5]&  2.35[-5]\\
 9&  1.62[-2]&   -2.86[-6]& -9.27[-5]&  1.79[-5]& -4.87[-5]&  1.81[-5]\\
\multicolumn{1}{c}{} &
\multicolumn{6}{c}{$\delta \langle 7S_{1/2} |h_W| nP_{1/2}\rangle$} \\
 6&  2.72[-2]&  -4.74[-6]& -1.53[-4]&  2.96[-5]& -8.06[-5]&  3.00[-5]\\
 7&  1.54[-3]&  -2.84[-6]& -8.96[-5]&  1.77[-5]& -4.83[-5]&  1.80[-5]\\
 8&  1.04[-3]&  -1.95[-6]& -6.06[-5]&  1.21[-5]& -3.31[-5]&  1.23[-5]\\
 9&  0.78[-3]&  -1.50[-6]& -4.56[-5]&  9.36[-6]& -2.55[-5]&  9.51[-6]\\
\end{tabular}
\end{table}

\begin{table}
\caption{ Summary of corrections to  PNC amplitude in $^{133}$Cs
due to Breit interaction and negative-energy states. Line $\delta H_{\rm W}$ lists contributions due to modifications
in the weak interaction matrix elements, and $\delta D$ due to corrections
in the electric dipole matrix elements.
See the Table~\ref{Tbl_Cs_weak} caption   for the explanation of 
columns.
The units are $10^{-11} i (-Q_{\rm W}/N)$ , and 
$x[y] = x \times 10^{y}$. \label{Tab_Epnc}}  
\begin{tabular}{lrrrr}
&
\multicolumn{1}{c}{$C_-$} & 
\multicolumn{1}{c}{$B^{(1)}_{\pm}$} & 
\multicolumn{1}{c}{$B^{(2)}_{\pm}$} & 
\multicolumn{1}{c}{$\delta E_{\mathrm PNC} $} \\
\hline           
$\delta H_{\rm W}$ &  0.0002 & 0.0042   &  0.0019    &  0.0063  \\
$\delta D$ & -2.6[-10]&-0.0008   &  1.3[-6]   & -0.0008 \\
Total        &  0.0002  & 0.0034   &  0.0019    &  0.0055 \\
\end{tabular}
\end{table}

\begin{table}
\caption{ Contributions to hyperfine-structure constants in MHz. 
Column ``Expt'' lists experimental values, where available, and
$\delta A$ gives the total of the contributions from negative-energy
states and Breit interaction.
See the Table~\ref{Tbl_Cs_weak} caption   for the explanation of other
columns. Notation $x[y]$ means $x \times 10^{y}$.
\label{Tbl_hfs}}
\begin{tabular}{lr@{.}lrrrrlr}
\multicolumn{1}{c}{state} & 
\multicolumn{2}{c}{Expt} & 
\multicolumn{1}{c}{$C_-$} & 
\multicolumn{1}{c}{$B^{(1)}_+$} & 
\multicolumn{1}{c}{$B^{(1)}_-$} & 
\multicolumn{1}{c}{$B^{(2)}_+$} & 
\multicolumn{1}{c}{$B^{(2)}_-$} &
\multicolumn{1}{c}{$\delta A$}  \\
\hline
\multicolumn{9}{c}{$^{133}{\mathrm Cs},\; \; g_I = 0.73789, \; \; R_m = 5.6748$ fm }   \\
$6S_{1/2}$& 2298&2   &   0.11    & -8.14  &  0.25 &  3.50 & -0.35 & -4.64\\
$7S_{1/2}$& 545&90(9)&   0.03    & -1.80  &  0.07 &  0.96 & -0.097& -0.83\\
$6P_{1/2}$& 291&89(9)&  -6.1[-4]& -1.58  &  0.25 &  0.73 & -0.27 & -0.87 \\
$7P_{1/2}$& 94&35    &  -2.2[-4]& -5.43  &  0.09 &  0.26& -0.098& -0.29\\[5pt]
\multicolumn{9}{c}{$^{211}{\mathrm Fr},\; \; g_I =0.888 , \; \; R_m = 6.71$ fm } \\
 $7S_{1/2}$&  8713&9(8)  & 0.07 & -66.7 & -0.08  & 19.8 & -0.54 &-47.4\\
 $8S_{1/2}$&  1912&5(1.3)& 0.02 & -12.8 & -0.02&  5.08  & -0.14 &-7.88\\
 $7P_{1/2}$&  1142&0(3)  & -5.6[-3]& -10.8&  1.23&  3.62& -0.95&-6.90\\
 $8P_{1/2}$&  362&91\tablenotemark[1]   & -2.0[-3]& -3.61&  0.44&  1.29& -0.34&-2.23\\
\end{tabular}
\tablenotetext[1]{All-order many-body calculations Ref.~\protect\cite{Safronova_AlkSDpT99}.}
\end{table}

\end{document}